\documentclass[10pt, journal, compsoc]{IEEEtran}

\ifCLASSOPTIONcompsoc
  \usepackage[nocompress]{cite}
\else
   \usepackage{cite}
\fi
\ifCLASSINFOpdf
    \usepackage[pdftex]{graphicx}
   \graphicspath{{. . /pdf/}{. . /jpeg/}}
   \DeclareGraphicsExtensions{. pdf, . jpeg, . png}
\else
   \usepackage[dvips]{graphicx}
   \DeclareGraphicsExtensions{. eps}
\fi
\usepackage{amsmath}
\usepackage{array}
\usepackage{float}

\usepackage{url}
\usepackage{multirow}
\usepackage[]{hyperref}
\usepackage{longtable}

\begin{document}

%
\title{Stacked Autoencoder Based\\ Multi-Omics Data Integration\\
for Cancer Survival Prediction }
%
%
%
%

\author{Xing~Wu, 
        Qiulian~Fang
\IEEEcompsocitemizethanks{
\IEEEcompsocthanksitem X. Wu is with the Department of mathematics and statistics, Central South University,  Changsha, HN 410083,  China. \\
E-mail:  wuxing98@csu.edu.cn
\IEEEcompsocthanksitem Q.  Fang is with the Department of mathematics and statistics, Central South University,  Changsha, HN 410083,  China. \\
E-mail:  qiulianfang@csu.edu.cn
}
\thanks{Manuscript received. . .  }}




\IEEEtitleabstractindextext{%
\begin{abstract}
Cancer survival prediction is important for developing personalized treatments and inducing disease-causing mechanisms.
Multi-omics data integration is attracting widespread interest in cancer research for providing information for understanding cancer progression at multiple genetic levels. Many works, however, are limited because of the high dimensionality and heterogeneity of multi-omics data.
In this paper, we propose a novel method to integrate multi-omics data for cancer survival prediction, called Stacked AutoEncoder-based Survival Prediction Neural Network (SAEsurv-net). In the cancer survival prediction for TCGA cases, SAEsurv-net addresses the curse of dimensionality with a two-stage dimensionality reduction strategy and handles multi-omics heterogeneity with a stacked autoencoder model. The two-stage dimensionality reduction strategy achieves a balance between computation complexity and information exploiting. The stacked autoencoder model removes most heterogeneities such as data’s type and size in the first group of autoencoders, and integrates multiple omics data in the second autoencoder.
The experiments show that SAEsurv-net outperforms models based on a single type of data as well as other state-of-the-art methods.
The source code can be found on the Github: \url{https://github.com/luckydoggy98/SAEsurv-net}.
\end{abstract}

}

\maketitle

\IEEEdisplaynontitleabstractindextext

%
\IEEEpeerreviewmaketitle

\ifCLASSOPTIONcompsoc
\IEEEraisesectionheading{\section{Introduction}\label{sec: introduction}}
\else

\section{Introduction}
\label{sec: introduction}
\fi
\IEEEPARstart{L}{ooking} back to the previous decade, feeding integrating multi-omics data and clinical data as input to models leads to a significant increase in accuracy of survival prediction, which facilitates understanding biological processes of disease and personalizing optimal therapies. 
Cancer is associated with abnormal cell growth, uncontrollable cell division and other tissues of the body through metastasis. The onset and progression of cancer can occur under the influence of complicated mechanisms and alterations at various levels including genome, proteome, transcriptome, and metabolome etc.\cite{Arjmand}. Therefore, integration strategies across multiple cellular function levels provide unprecedented opportunities to understand the underlying biological mechanisms of cancer\cite{Menyh}.

The term "multi-omics" refers to the study of the genome, transcriptome, epigenome, proteome, exposome, and microbiome\cite{kang2022}. Specifically, genomics is the study of identifying genetic variants associated with disease at the genome scale. Transcriptomics is the study of RNA expression from specific tissues, which provides information about cell- and tissue-specific gene expression. Epiegnomics are concerned with the genome-wide characterization of DNA methylation or post-translational histone modifications capable of imposing stable and heritable changes in a gene without changing the DNA sequence. Metabolomics provides a snapshot of an organism's or tissue's metabolic state. Proteomics is the qualitative and quantitive study of the proteome. Exposomes are non-genetic disease drivers that represent the totality of environmental exposure throughout a person's life. Microbiomes are microorganisms influencing the physiology of an individual, such as bacteria, viruses, and fungi\cite{kang2022}. 
High-throughput experimental technologies, including next-generation sequencing, microarrays, and RNA-sequencing technologies provide exponentially increased amount of available omics data \cite{rigden2016}.
Combining properly these types of omics data as predictor variables allows to comprehensively model biological variation at various levels of regulation\cite{ritchie2015}. 

However, challenges such as the curse of dimensionality and data heterogeneity arise when integrating multi-omics data.
Because of the large number of measured objects and the small number of patients, the curse of dimensionality occurs naturally in omics data. After integrating, the situation worsens because the number of features to be analyzed grows while the number of samples remains constant. This often arises the risk of overfitting\cite{domingos2012}. 
The heterogeneity of multi-omics data is on account of different number of features, mismatched data distributions, scalings and modalities between various omics\cite{mirza2019}. These discrepancies between omics can hinder their integration.  

Recent researches mainly focus on addressing these challenges. 
Dimensionality reduction techniques are used in two ways: feature selection, which identifies a relevant subset of the original features, and feature extraction, which projects data from a high-dimensional space to a lower-dimensional space\cite{wang2016}. 
Feature selection methods can be classified into three main types: (1) filter-based, (2) wrapper-based, (3) embedded methods. Filter-based methods are independent of any predictive model and implement statistical analysis to select a subset of relevant features based on correlation (e.g., CFS\cite{hall1999}), distance (e.g., ReliefF\cite{kononenko1994}), information gain(e.g., mRMR\cite{ding2005}) and statistical test\cite{fu2018}. Wrapper-based methods try to search the best subset of features by applying predictive model repeatedly on different set of features, and then keep the best performing subsets (e.g., RFE\cite{guyon2002}), but this strategy is limited by its computing efficiency especially when the dataset is large. Embedded methods are algorithms with feature selection built directly in the predictive models (e.g., tree based feature importance\cite{scornet2020}, LASSO). 
The most widely used feature extraction method is Principal Component Analysis (PCA)\cite{ringner2008}, which transforms high-dimensional features to linearly uncorrelated new features named principal components.  
However, the transformation in PCA is a linear combination, so that PCA cannot handle non-linear trends in data. 
Autoencoder is a nonlinear feature transformation technique with multi-layer neural networks structures to extract latent features in a bottleneck layer, which can be employed to reduce dimensionality by restricting the number of bottleneck layer nodes\cite{kang2022}. 
Both feature selection and feature extraction methods can be used in the data integration analysis, either separately on each omics dataset before integrating or on concatenated multi-omics datasets. Selecting or extracting features from separate datasets before integrating reduces the computation complexity of the subsequent analysis, but it ignores the relationship between the datasets and commonly results in unwanted redundancy and suboptimal results\cite{tang2021}. When dealing with concatenated datasets, it is possible to capture information shared across different datasets and discover more revelatory features that single-omics would miss. However, it must deal with a high-dimensional input data matrix. In order to achieve a balance between computation efficiency and information exploitation quality, the dimensionality reduction is expected to combine with the data integration to be a syncretic process. 

M. Picard \emph{et al.} classify integration methods into five types: early (concatenation-based), mixed (transformation-based), late (model-based), intermediate and hierarchical\cite{picard2021}. 
The early integration is based on concatenating every dataset into a single large matrix\cite{Yousefi2017}. Although simple, this process results in a more complex, noisier and higher dimensional matrix, and ignores the specific data distribution of each omics\cite{picard2021}. 
The mixed strategy addresses the shortcomings of the early integration by independently transforming each dataset into a simpler representation before concatenating, which can be less dimensional and less noisy\cite{picard2021}. Commonly used transformation methods are kernel-based, graph-based and neural network-based. Artificial neural networks can capture nonlinear and hierarchical new features from the hidden layers due to their multi-layered structures and nonlinear activation functions, which process can be considered as successive feature extraction. Z Huang \emph{et al.} propose an algorithm called SALMON to represent mRNA-seq and miRNA-seq data with neural networks respectively. Then the output layers of these networks are fed into a full connected layer together to predict death risk\cite{Huang2019}. Some neural networks have architectures specialized in learning meaningful representation, such as autoencoder. New representations for each dataset can be extracted by type-specific encoding sub-networks\cite{sharifi2019,karim2019}. 
Intermediate integration can be described as any methods capable of jointly integrating the multi-omics datasets without prior transformation\cite{picard2021}. For example, K Chaudhary \emph{et al.} feed stacked three omics data matrices into an autoencoder to extract new integrated features\cite{chaudhary2018}. H Torkey \emph{et al.} train autoencoder based on concatenated data and take the decoder layer as the new features\cite{torkey2021}. Similar to the early integration, the large input matrix in intermediate integrating process is also difficult to exploit. 
Late integration means to apply predictive models separately on each dataset and then combines their predictions\cite{picard2021}. Sun \emph{et al.} use a set of deep neural networks to extract information from different datasets respectively, then conduct a score-level fusion for the results predicted by each DNN model\cite{sun2018}. The limitation of this strategy is that it cannot capture inter-omics interactions since it only combines predicted results\cite{picard2021}. 
Hierarchical integration strategy bases the integration on inclusion of the prior knowledge of regulatory relationships between different datasets, which means external information from interaction database and scientific literature is used\cite{picard2021}. For example, J Hao \emph{et al.} propose a neural network named MiNet, which has a structure following a biological system by utilizing prior knowledge of biology pathways\cite{hao2019}. On the one hand, this strategy improves model interpretability; on the other hand, its reliance on prior knowledge limits its ability to investigate new biological mechanisms.

Inspired by successful use of integrating strategies in multi-omics data analysis, particularly the use of the neural networks for dimensionality reduction and data integration, we propose a novel approach to integrate multi-omics data for survival analysis called Stacked AutoEncoder based survival prediction neural network (SAEsurv-net).  
The main contributions of this paper can be summarized as follows: 
(1) achieving a balance between computation complexity and information exploiting by applying a two-stage dimensionality reduction strategy that first removes features irrelevant to the survival outcome for each data type and then combines the remaining features from various data types;
(2) handling heterogeneity of multiple data types by introducing a stacked autoencoder model, in which most heterogeneities such as data's type and size are removed in the first group of autoencoders;
(3) comparing the performance of SAEsurv-net with neural network models based on a single data type and other previous state-of-the-art models. 

The rest of the paper is organized as follows: Section 2 elaborates the overall approach, including data collection and preprocessing, as well as the construction of the SAEsurv-net. Section 3 shows the experimental settings and results. Section 4 summarizes the research, identifies potential limitations, and suggests future directions. 

\section{Methods And Materials}

\begin{table*}[!h]
\centering
\caption{The characteristics of the data}
\begin{tabular}{cccccc}
\hline
Disease & Data Category & Number of Samples & Number of Features & Number of Features after Preprocess & Number of Features after SAE \\ \hline
\multirow{3}{*}{GBM}  & gene    & \multirow{3}{*}{156}  & 60484 & 4080 & \multirow{2}{*}{totally 400}\\ 
                      & cnv        &       & 19730         & 238  &     \\                
                      & clinical   &    & 3     & 2    &   2  \\                    
\multirow{3}{*}{OV}   & gene     & \multirow{3}{*}{373}  & 60484 & 3503 & \multirow{2}{*}{totally 365} \\
                      & cnv      &                       & 19730 & 1075 &                      \\
                      & clinical &                       & 5     & 2    &         2             \\
\multirow{3}{*}{BRCA} & gene     & \multirow{3}{*}{1040} & 60484 & 5313 & \multirow{2}{*}{totally 595} \\
                      & cnv      &                       & 19730 & 1042 &                      \\
                      & clinical &                       & 9     & 3    &         3             \\ \hline
\end{tabular}
\label{feature}
\end{table*}

\subsection{Materials}
To evaluate the performance of SAEsurv-net, we download glioblastoma multiforme (GBM), ovarian serous cystadenocarcinoma (OV), and breast invasive carcinoma (BRCA) datasets from The Cancer Genome Atlas (TCGA) project through UCSC Xena\cite{Goldman}. 
Each dataset contains three types of information: gene expression, copy number variations (CNV), and clinical information. 
Gene expression features are taken as FPKM values from the Illumina platform. CNV features are derived from Affymetrix Genome Wide Human SNP 6.0 platform. 
Characteristics for each dataset are shown in Table \ref{feature}. Each gene expression dataset has 60484 features, each CNV dataset has 19730 features, and each clinical dataset of GBM, OV, and BRCA has 3, 5, and 9 features, respectively. Features involved in each clinical dataset are listed in Appendix A.
After removing the patients who do not have all the three types of information simultaneously, we get 156, 373, 1040 samples in GBM, OV, BRCA data respectively. 

\begin{figure*}[!t]
\centering
\includegraphics[width=0.8\textwidth]{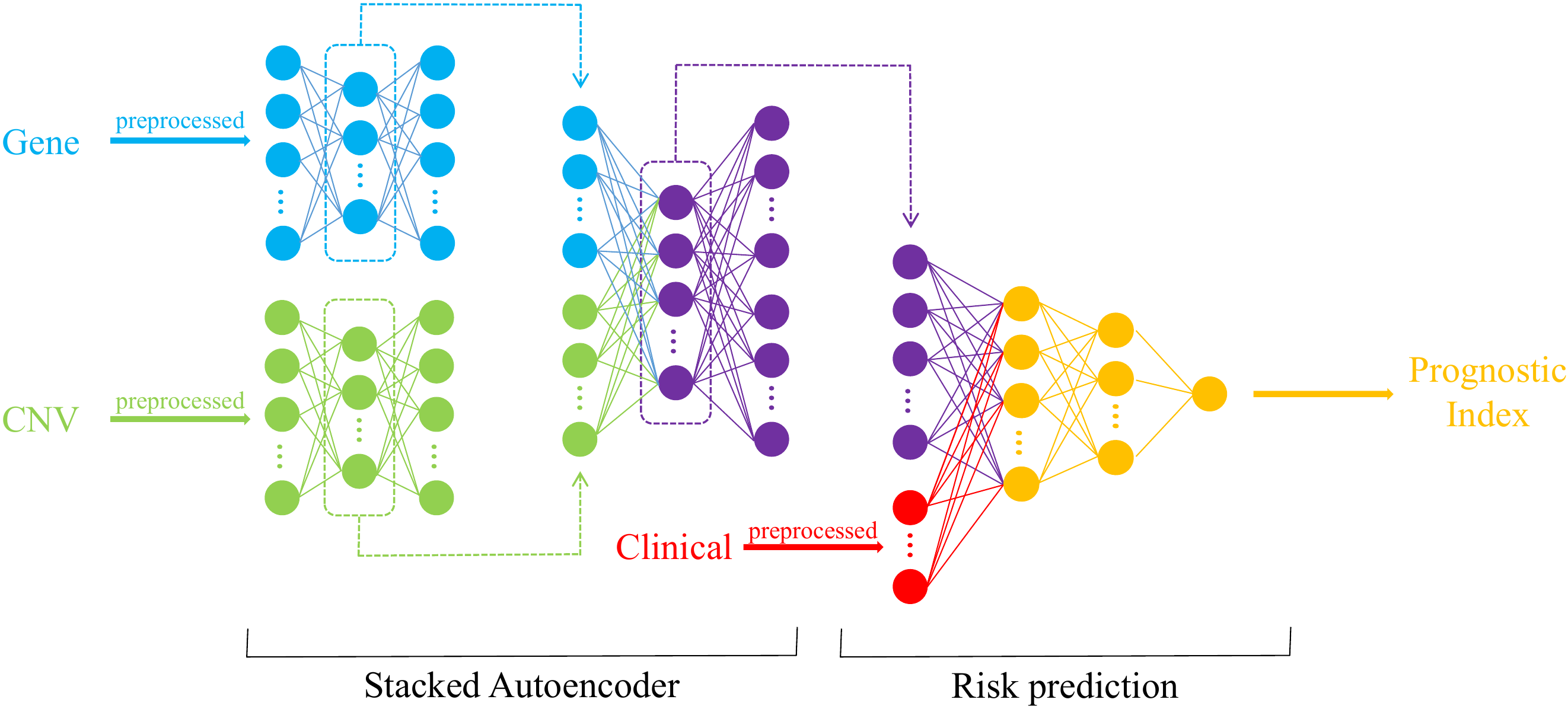}
\DeclareGraphicsExtensions{.pdf}
\caption{Architecture of SAEsuvr-net. In the first place, gene expression (Gene), copy number variations (CNV) and clinical data are prepocessed. In this process, all features are normalized and redundant features are removed. Next, the preprocessed gene expression and CNV vectors are fed into two one-hidden-layer autoencoder models respectively. From these models, the low-dimensional representations for corresponding data can be obtained. Then, the representations are concatenated and fed into another one-hidden-layer autoencoder. After that, a cross-omics representation is obtained. Subsequently, the cross-omics representation is combined with clinical vector and inputted to a risk prediction network with two hidden layers. The final outcome of the risk prediction network is Prognostic Index (PI), indicating the death risk of samples.}
\label{process}
\end{figure*}

\subsection{Preprocess}
In the first place, We normalize all datasets and remove redundant features.
In gene expression data, features are applied $\log 2(1+x)$ transformation and zero-mean unit-variance normalization. In clinical data, categorical features are transformed into one-hot codes and numerical features are normalized to zero-mean unit-variance, while the missing values are imputed by mode. 

To avoid overfitting, we adopt a two-stage dimensionality reduction strategy. 
In the first stage, we remove features with zero variance and features with the same value for each type of data. After that, the log-rank test is applied to each feature, and features with the p-values greater than 0.05 are removed. Before the log-rank test, features in gene expression are discretized into three categories: under-expression (-1), over-expression (1), and baseline (0), and numerical values are used after the test. The results of this stage are shown in Table \ref{feature}. For all diseases, the decrease of the numbers of features in gene expression and CNV is greater than $90\%$. 
In the second stage, we apply a stacked autoencoder model to extract and integrate features from multiple omics data.  

\subsection{A Stacked Autoencoder Model for Feature Extraction and Integration}
After the first stage of dimensionality reduction, we obtain a subset of original features for each single dataset, however, the number of features is still too large when compared with the sample size. 
In the second stage, we create a lower-dimensional representation for multiple omics with a stack of autoencoder models. This procedure can be divided into two steps. The first step is to train an autoencoder on each dataset. By limiting the number of nodes in the bottleneck layer, a lower-dimensional representation of each dataset can be obtained. In the new representations, most heterogeneities such as data’s type and size are removed. In the second step, representations corresponding to multiple omics data are concatenated and fed into another autoencoder. This step simultaneously reduces dimensionality and integrates data. Consequently, a cross-omics representation is obtained.

As shown in Fig. \ref{process}, the preprocessed gene expression vector $x_\mathrm{gene}$ and CNV vector $x_\mathrm{cnv}$ are fed into two autoencoder models, respectively.
Each autoencoder is a neural network that is trained to attempt to copy its input $x$ to its output $\hat{x}$\cite{Goodfellow2016}. The network consists of two parts: an encoder that represents the input and a decoder that produces a reconstruction. The most salient features of the training data will be captured by constraining the bottleneck layer $h$ to have a smaller dimension than $x$. Let $h_\mathrm{gene}$ and $h_\mathrm{cnv}$ denote the hidden representations of the two one-hidden-layer autoencoder models respectively:

\begin{equation}
    \label{hgene}
    h_{\mathrm{gene}}=f_{1,g}(W_{g} x_\mathrm{gene}+b_{g})
\end{equation}
\begin{equation}
    \label{hcnv}
    h_\mathrm{cnv}=f_{1,c}(W_{c} x_\mathrm{cnv}+b_{c})
\end{equation}
And the reconstruction is:
\begin{equation}
    \label{ggene}
    \hat x_\mathrm{gene}=g_{1,g}(W_{g}^{\prime}h_\mathrm{gene}+b_{g}^{\prime})
\end{equation}
\begin{equation}
    \label{gcnv}
    \hat x_\mathrm{cnv}=g_{1,c}(W_{c}^{\prime}h_\mathrm{cnv}+b_{c}^{\prime})
\end{equation}
Then the two hidden representations are concatenated as $[h_\mathrm{gene};h_\mathrm{cnv}]$ and inputted to another autoencoder, which extracts the cross-omics representation $h_\mathrm{con}$ of multiple omics data:
\begin{equation}
    \label{hcon}
    h_\mathrm{con}=f_{2}(W_\mathrm{con}[h_\mathrm{gene};h_\mathrm{cnv}]+b_\mathrm{con})
\end{equation}
And the reconstruction is:
\begin{equation}
    \label{gcon}
    \hat x_\mathrm{con}=g_{2}(W_\mathrm{con}^{\prime}h_\mathrm{con}+b_\mathrm{con}^{\prime}))
\end{equation}
This two-step data extraction process based on a stack of autoencoder models is a so-called Stacked AutoEncoder (SAE) model\cite{wang2019}. SAE is defined as a stacking of multiple autoencoder models to form a deep structure. In SAE, the hidden layer of the $k$th autoencoder is fed as the input into the $(k+1)$th autoencoder.

We define regularized mean squared error (MSE) loss as the objective function for each autoencoder:
\begin{equation}
    \label{aeg}
    \begin{aligned}
    L_{g}(W_{g},b_{g},W_{g}^{\prime},b_{g}^{\prime})= &
    \sum_{i=1}^{n}{\|\hat x_\mathrm{gene}^{i}-x_\mathrm{gene}^{i}\|_{2}^{2}}\\
    &+\lambda_{g}(\| W_{g} \|_{1}+\| W_{g}^{\prime} \|_{1})\\
    &+\alpha_{g}(\| W_{g} \|_{2}^{2}+\| W_{g}^{\prime} \|_{2}^{2})
    \end{aligned}
\end{equation}
\begin{equation}
    \label{aec}
    \begin{aligned}
    L_{c}(W_{c},b_{c},W_{c}^{\prime},b_{c}^{\prime})= &
    \sum_{i=1}^{n}{\|\hat x_\mathrm{cnv}^{i}-x_\mathrm{cnv}^{i}\|_{2}^{2}}\\
    &+\lambda_{c}(\| W_{c} \|_{1}+\| W_{c}^{\prime} \|_{1})\\
    &+\alpha_{c}(\| W_{c} \|_{2}^{2}+\| W_{c}^{\prime} \|_{2}^{2})
    \end{aligned}
\end{equation}
\begin{equation}
    \label{aecon}
    \begin{aligned}
    L_\mathrm{con}(W_\mathrm{con},b_\mathrm{con},W_\mathrm{con}^{\prime},b_\mathrm{con}^{\prime})&= 
    \sum_{i=1}^{n}{\|\hat x_\mathrm{con}^{i}-[h_\mathrm{gene}^{i};h_\mathrm{cnv}^{i}]\|_{2}^{2}}\\
    &+\lambda_\mathrm{con}(\| W_\mathrm{con} \|_{1}+\| W_\mathrm{con}^{\prime} \|_{1})\\
    &+\alpha_\mathrm{con}(\| W_\mathrm{con} \|_{2}^{2}+\| W_\mathrm{con}^{\prime} \|_{2}^{2})
    \end{aligned}
\end{equation}
where $\| \cdot \|_{1}$ and $\| \cdot \|_{2}^{2}$ denote $L_{1}$ and $L_{2}$ norm, and $i$ denotes the $i$th feature. The weight parameters $W_{g}$, $W_{g}^{\prime}$, $W_{c}$, $W_{c}^{\prime}$, $W_\mathrm{con}$, $W_\mathrm{con}^{\prime}$ and bias parameters $b_{g}$, $b_{g}^{\prime}$, $b_{c}$, $b_{c}^{\prime}$, $b_\mathrm{con}$, $b_\mathrm{con}^{\prime}$ in neural networks can be obtained by minimizing the loss functions respectively, the hyperparameters such as activation function, numbers of nodes, $\alpha_{g}$, $\lambda_{g}$, etc., can be obtained by optimizing algorithm described in Section 3.1.

\subsection{A Risk Prediction Neural Network Model for Survival Analysis}
Once the stack of autoencoder models has been built, its highest level output representation $h_\mathrm{con}$ can be used in the following prediction.
As shown in Fig. \ref{process}, the vector $x_\mathrm{in}$, which is a concatenation of $h_\mathrm{con}$ and clinical vector $x_\mathrm{cli}$, is inputted to a risk prediction network with two hidden layers. The final outcome of the risk prediction network is Prognostic Index (PI), which is a linear combination of the neurons in the previous layer. PI is introduced to the hazard function $h(t|x_\mathrm{in})$ for the Cox proportional hazards model as an estimate of risk  function\cite{hao2019}:
\begin{equation}
    h(t|x_\mathrm{in})= 
    h_{0}(t)\mbox{exp}(\mbox{PI})
\end{equation}
where $x_\mathrm{in}=[h_\mathrm{con};x_\mathrm{cli}]$, and $h_{0}(t)$ is the baseline hazard function which can be an arbitrary non-negative function of time. This hazard function shows the death rate at time $t$ conditional on survival until time $t$ or later. We can say that samples with higher PI have higher death risk. The loss function of the risk prediction network is the regularized negative $\mathrm{log}$ partial likelihood:
\begin{equation}
    \begin{aligned}
        L_{r}(\Theta_{r})= & -\sum_{i:\delta_{i}=1}
        \left(\mbox{PI}_{\Theta_{r}}\left(x_{\mathrm{in},i}\right)
        -\log \sum_{j \in \Re\left(T_{i}\right)} e^{\mbox{PI}_{\Theta_{r}}\left(x_{\mathrm{in},j}\right)}\right)\\
        & +\lambda_{r}\| W_{r} \|_{1}+\alpha_{r}\| W_{r} \|_{2}^{2}
    \end{aligned} 
\end{equation}
where $T_{i}$, $E_{i}$, $\Re\left(T_{i}\right)$ denotes the death time, censored status and risk set for the $i$th death, respectively. $\delta_{i}=1$ means the death time of the $i$th sample is observed, and $\delta_{i}=0$ means the death time of the $i$th sample is censored.
Then $\Theta_{r}=\{W_{r},b_{r}\}$ can be obtained by minimizing the loss function, and the hyperparameters such as activation function, numbers of nodes, $\alpha_{r}$, $\lambda_{r}$, etc., can be obtained by the optimizing algorithm described in Section 3.1.

\begin{figure*}[!t]
\centering
\includegraphics[width=0.98\textwidth]{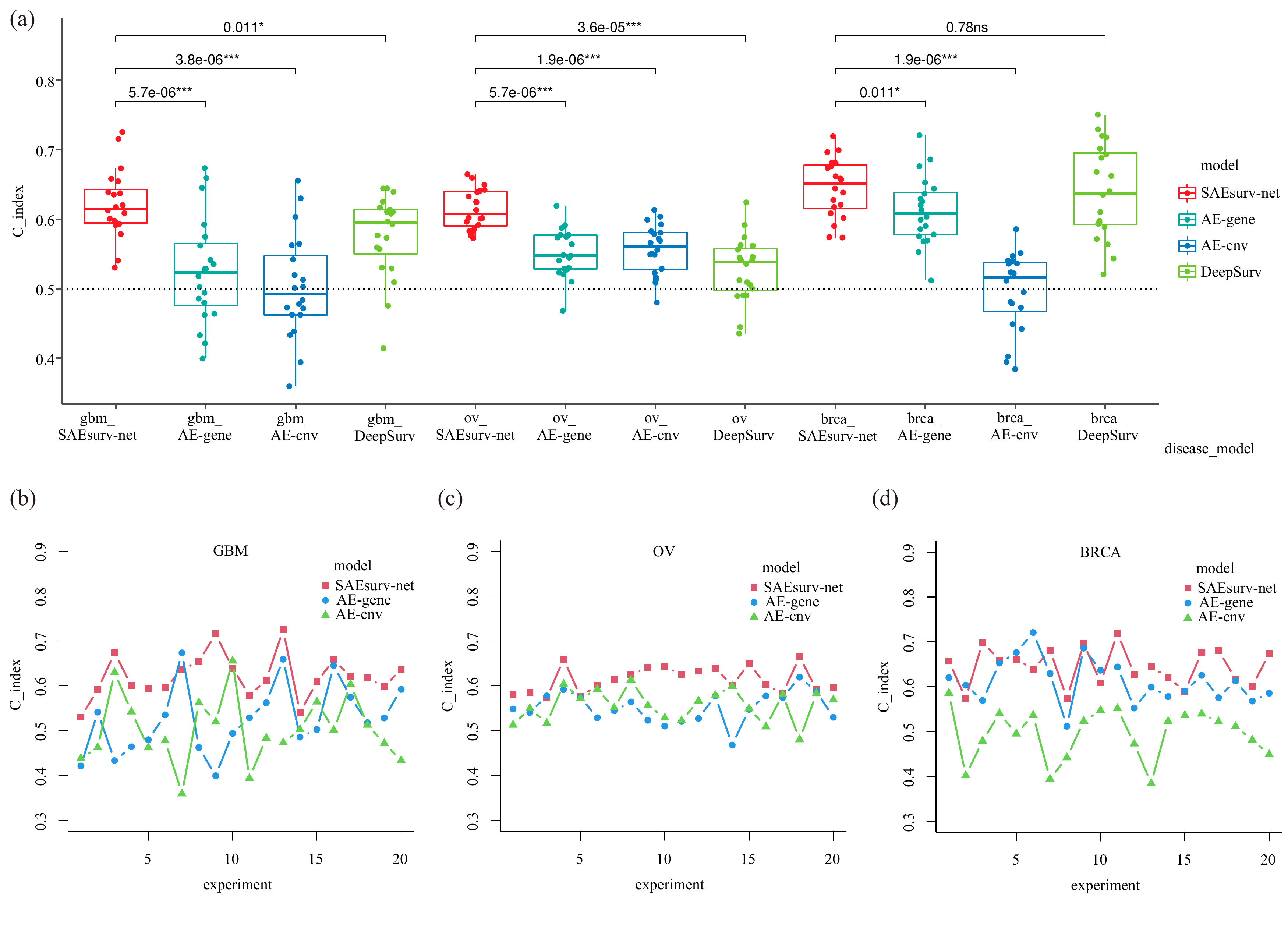}
\DeclareGraphicsExtensions{.pdf}
\caption{C-index in 20 experiments. 
(a) Box plots of C-index in 20 experiments. Wilcoxon test results are represented by p-values and symbol *. *** denotes a p-value lower than 0.001, ** denotes a p-value bigger than 0.001 but lower than 0.01, * denotes a p-value bigger than 0.01 but lower than 0.05, ns denotes a p-value bigger than 0.05.
(b),(c) and (d) C-index in each experiment for GBM, OV and BRCA.}
\label{box}
\end{figure*}

\section{Results}

\subsection{Experimental Design}
The performance of SAEsurv-net is assessed by comparing it with networks based on a single data type and current state-of-the-art methods, such as DeepSurv\cite{katzman2018}. 
We consider C-index\cite{Uno2011} as the measurement of the predictive performance. It is defined as the ratio of the concordant predicted pairs to the total comparable pairs: 
\begin{equation}
    c= \frac{1}{\mathrm{num}} \sum_{i: \delta _{i}=1} \sum_{j: y_{i}<y_{j}}I\left[\mbox{PI}_{i}>\mbox{PI}_{j}\right]
\end{equation}
where $\mathrm{num}$ denotes the total number of comparable pairs, $I[\cdot]$ is the indicator function. $\delta _{i}$, $y_{i}$ and $\mbox{PI}_{i}$ denote the censored status, observed survival time and predicted death risk of the $i$th sample.

We repeat the evaluation 20 times for the reproducibility of the model performance.
On each experiment, data is randomly split into two subsets: training ($80\%$) and test data ($20\%$) by stratified sampling with censored status. We train every model with training data, and apply a 5-fold cross validation to find the optimal hyperparameters.
Once the model was well-trained, the test data is used to evaluate the predictive performance.

The neural networks are implemented based on TensorFlow 1.15.2 and Keras 2.8.0 deep learning library; the data preprocessing and train-test split are implemented with scikit-learn \cite{sklearn_api}; the log rank test and C-index computation are performed with lifelines \cite{Davidson-Pilon2019}; the plots of results are accomplished through ggplot2\cite{ggplot2} and ggpubr with Rstudio. More details of SAEsurv-net can be found at \url{https://github.com/luckydoggy98/SAEsurv-net}. 
In oder to acheive a nearly fair comparison, we train DeepSurv by following the implementation codes provided by the autors' GitHub \url{https://github.com/jaredleekatzman/DeepSurv}. 

We perform the hyperparameter optimization through Bayesian optimization implemented with BayesianOptimization\cite{bayesopt} and NNI\cite{Microsoft_Neural_Network_Intelligence_2021}. The latter is an open source AutoML toolkit for hyperparameter optimization, neural architecture search, model compression and feature engineering. For autoencoder models, the configuration with the least validation mean squared error will be chosen, for risk prediction network, the configuration with the largest validation C-index will be chosen.
The detailed hyperparameters used in SAEsurv-net are described in Appendix B.

\begin{figure*}[!t]
\centering
\includegraphics[width=0.98\textwidth]{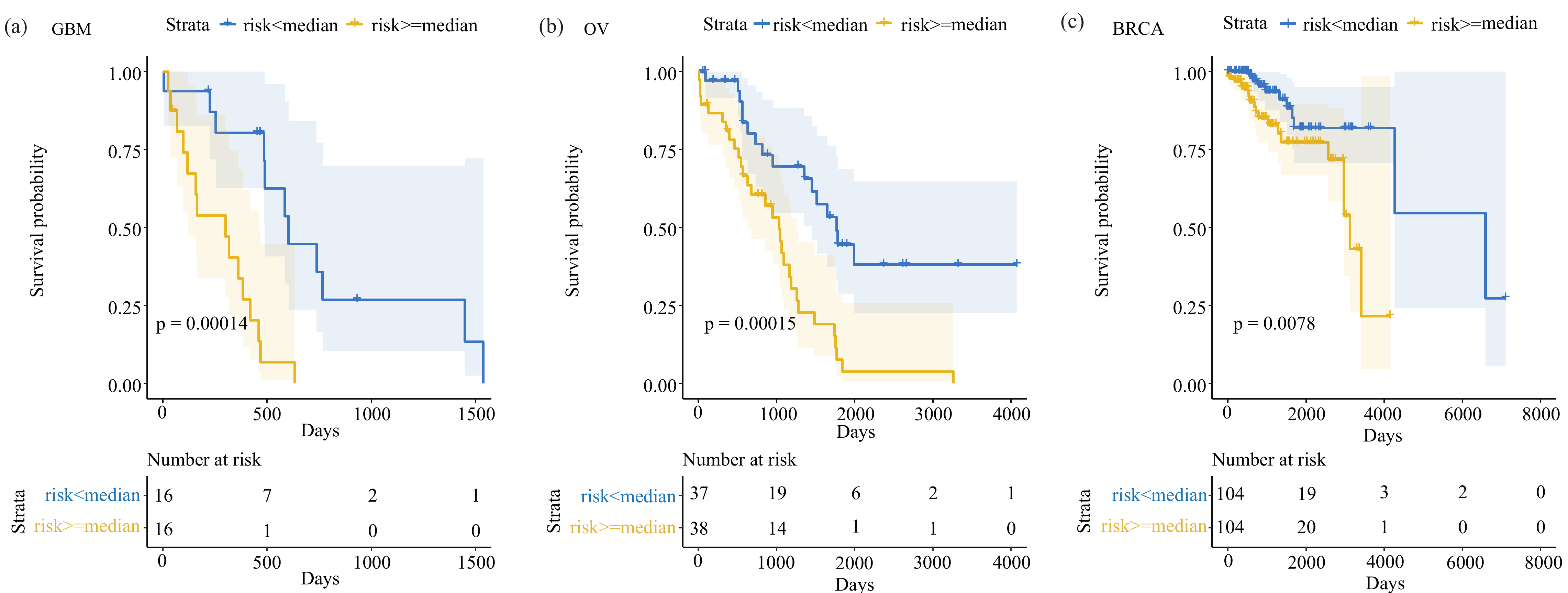}
\DeclareGraphicsExtensions{.pdf}
\caption{Kaplan-Meier plots for risk predicted by SAEsurv-net in (a) GBM , (b) OV, and (c) BRCA dataset. For each test data, samples with the predicted risk lower than the overall median of predicted risk are assigned to the low-risk group, otherwise they are assigned to the high-risk group. Tables below present the number of samples belonging to the low- and high-risk groups at certain times.}
\label{km}
\end{figure*}

\subsection{Comparison of SAEsurv-net with Single-Omics Data Based Models}
To illustrate the effectiveness of integration strategy, we  compare the performance of SAEsurv-net with neural networks based on a single type of data. For simplicity, we denote networks only based on gene expression or CNV data as AE-gene and AE-cnv, respectively. In AE-gene or AE-cnv, preprocessed gene expression or CNV data is fed into a one-hidden-layer autoencoder, as in SAEsurv-net. Then the representations extracted are fed into a two-hidden-layer risk prediction network. There is no integrating step in AE-gene nor AE-cnv. The loss functions and hyperparameter optimization methods in AE-gene, AE-cnv, and SAEsurv-net are the same.

Fig. \ref{box}(a) shows the distribution of C-index of the experiments. The mean values of C-index are listed in Table  \ref{mean}. For each disease, SAEsurv-net achieves the highest mean (0.6212, 0.6141, and 0.6451 respectively) and relatively low variance of C-index, compared with AE-gene (0.5250, 0.5516, and 0.6119 respectively) and AE-cnv (0.5025, 0.5566, and 0.4958 respectively). * in Fig. \ref{box}(a) denotes the Wilcoxon rank-sum test results for two groups of C-index. *** denotes a p-value less than 0.001, ** denotes a p-value greater than 0.001 but less than 0.01, * denotes a p-value greater than 0.01 but less than 0.05, ns denotes a p-value greater than 0.05. We can see that the outperformance of SAEsurv-net against AE-gene and AE-cnv is statistically significant (p-value$<$0.05).
As shown in Fig. \ref{box}(b)(c)(d), SAEsurv-net has the highest C-index in almost every experiment. It means that the performance improvement of SAEsurv-net is consistent and does not rely on data splitting. 

\begin{table}[!h]
\centering
\caption{Mean values of C-index over 20 experiments}
\resizebox{\linewidth}{!}{ 
\begin{tabular*}{\hsize}{@{}@{\extracolsep{\fill}}ccc@{}}
\hline
Disease                         &  Model   & C-index mean \\ \hline
\multirow{4}{*}{GBM}            & AE-gene   & 0.5250  \\
                                & AE-cnv    & 0.5025    \\
                                & DeepSurv  & 0.5763  \\
                                & \textbf{SAEsurv-net} & \textbf{0.6212} \\
\multirow{4}{*}{OV}             & AE-gene   & 0.5516  \\
                                & AE-cnv    & 0.5566    \\
                                & DeepSurv  & 0.5277  \\
                                & \textbf{SAEsurv-net} & \textbf{0.6141} \\
\multirow{4}{*}{BRCA}           & AE-gene   & 0.6119  \\
                                & AE-cnv    & 0.4958    \\
                                & DeepSurv  & 0.6395  \\
                                & \textbf{SAEsurv-net} & \textbf{0.6451} \\\hline
\end{tabular*}
}
\label{mean}
\end{table}

\subsection{Comparison of SAEsurv-net with Other Methods}
To demonstrate the outperformance of SAEsurv-net against other integration strategies, We compare the performance of SAEsurv-net with DeepSurv. As shown in Table \ref{mean} and Fig. \ref{box}(a), for GBM and OV data, SAEsurv-net has higher C-index means (0.6212 and 0.6141) than DeepSurv with the p-values less than 0.05. It means that SAEsurv-net significantly outperforms DeepSurv on these two datasets. For BRCA data, the p-value of test between SAEsurv-net and DeepSurv is greater than 0.05, which means there is no significant improvement of SAEsurv-net against DeepSurv. However, SAEsurv-net still holds a higher mean (0.6451) and lower variance of C-index, indicating that it performs more robustly than DeepSurv at the same accuracy level.

\subsection{Visualization of risk prediction with Kaplan-Meier plot}
We perform the Kaplan-Meier plot with log-rank test to visualize the risk prediction of SAEsurv-net in one experiment. For a specific disease, samples in the test set are grouped into low-risk and high-risk groups according to their predicted risk. Specifically, samples with predicted risk lower than the overall median of predicted risk are assigned to the low-risk group, otherwise they are assigned to the high-risk group.
Fig. \ref{km} shows the Kaplan-Meier plots of the three diseases. We can see the distinct separation of the two groups with significant log-rank test results (p-value$ < $0.01).

\section{Discussion and Conclusion}
Cancer survival prediction is important for developing personalized treatments and inducing disease-causing mechanisms. In this study, we established a novel neural network named SAEsurv-net to integrate multi-omics data and predict death risk of patients with cancer. SAEsurv-net addresses the curse of dimensionality with a two-stage dimensionality reduction strategy. In the first stage, irrelevant features in all datasets are removed. In the second stage, the remaining features are transformed and integrated into a lower-dimensional representation by a stack of autoencoder models. For details, SAE extracts type-specific representations for each dataset from the first autoencoder, then extracts cross-omics representation from the second autoencoder. Most heterogeneities such as data’s type and size are removed through the first group of autoencoders. So SAEsurv-net also handles the heterogeneity between different types of data. Compared with integration strategies that either concatenating data matrices at the start or combining prediction results at the end, SAEsurv-net takes a middle ground approach. Experimental results of SAEsurv-net for cancer survival prediction are promising, achieving a higher mean and lower variance of C-index when compared to the neural networks based on a single data type and other state-of-the-art models. The evaluation results demonstrate the effectiveness of data integration strategy and advantages of SAEsurv-net. 

However, there are still some areas to be explored. 
Firstly, the sample size is still small when compared with the number of model parameters. Considering about the difficulty and cost of obtaining a large amount of data, methods should be developed to maximize the use of available data. 
Transfer learning is focusing on transferring the knowledge from some source tasks to a target task when the latter has fewer high-quality training data\cite{pan2009}. 
Existing multi-omics studies transfer knowledge by combining datasets from different diseases\cite{Yousefi}, or by finetuning pretrained neural networks built on unlabeled dataset using labeled dataset\cite{sevakula2018}.
However, involving unfamiliar instances may introduce noise and proper instance selection principles need to be studied.
Additionally, the lack of interpretability of deep neural networks hinders their application to real situations, and interpretable models should be established. Tree-based models and graph-based models\cite{xu2019} can provide explicit decision-making processes, which can generate intelligible rules. Introducing biological knowledge is also helpful to enhance the interpretability and reduce the overfitting risk of models\cite{hao2019}.


%
\newpage
\appendices

\section{Features Used in Clinical Data}

\begin{table}[!h]
\caption{Features Used in Clinical Data}
\label{clinical}
\resizebox{\linewidth}{!}{ 
\begin{tabular}{p{0.8cm}p{3.5cm}p{3.8cm}}

\hline
Disease     &  Feature name   & Description \\ \hline

\multirow{5}{*}{GBM}  & age at initial pathologic diagnosis 
                        & age at which a disease was first diagnosed\\
                      & radiation.therapy    
                        &  the patient's history of radiation therapy \\
                      & gender.demographic  
                        & gender  \\
\multirow{16}{*}{OV}   & age at initial pathologic diagnosis 
                        & age at which a disease was first diagnosed\\
                      & clinical.stage    
                        & stage group determined on the tumor, regional node and metastases and by grouping cases with similar prognosis for cancer \\
                      & neoplasm.histologic.grade  
                        & the degree of abnormality of cancer cells\\
                      & person.neoplasm.cancer.status 
                        & the state or condition of an individual’s neoplasm \\
                      & race.demographic  
                        & race based on the Office of Management and Budget  categories \\
\multirow{31}{*}{BRCA} & age at initial pathologic diagnosis 
                        & age at which a disease was first diagnosed\\
                      & history of neoadjuvant treatment 
                        & the patient’s history of neoadjuvant treatment and the kind of treatment given prior to resection of the tumor \\
                      & menopause.status  
                        & the status of a woman’s menopause \\
                      & pathologic.M  
                        & code to represent the defined absence or presence of distant spread or metastases (M) to locations  \\
                      & pathologic.N  
                        & codes that represent the stage of cancer based on the nodes present (N stage) \\
                      & pathologic.T  
                        & code of pathological T (primary tumor) to define the size or contiguous extension of the primary tumor (T)  \\
                      & ethnicity.demographic  
                        & ethnicity based on the Office of Management and Budget categories\\
                      & race.demographic  
                        & race based on the Office of Management and Budget categories\\
                      & person.neoplasm.cancer.status 
                        & the state of an individual’s neoplasm \\\hline
\end{tabular}
}

\end{table}

\section{Detailed Parameter Configurations of Models}

\begin{table}[!h]
\centering
\caption{Parameter configurations of the autoencoder on gene expression data}
\begin{tabular}{cccc}
\hline
\multirow{2}{*}{hyperparameter} & \multicolumn{3}{c}{value}   \\
                                & GBM      & OV     & BRCA    \\ \hline
dropout rate                    & 0.4567   & 0.3683 & 0.1694  \\
bottleneck units                & 290      & 450    & 460     \\
optimizer                       & adam     & nadam  & nadam   \\
activation function             & softsign & tanh   & tanh    \\
$\lambda_{g}$                  & 0        & 0      & 0       \\
$\alpha_{g}$                  & 0        & 0      & 0.00001 \\
learning rate                   & 0.001    & 0.001  & 0.0001  \\
training epoch                  & 50       & 80     & 100     \\
mini-batch size                 & 30       & 100    & 60     \\ \hline
\end{tabular}
\end{table}

\begin{table}[!h]
\centering
\caption{Parameter configurations of the autoencoder on CNV data}
\begin{tabular}{cccc}
\hline
\multirow{2}{*}{hyperparameter} & \multicolumn{3}{c}{value}   \\
                                & GBM      & OV     & BRCA    \\ \hline
dropout rate                    & 0.0220   & 0.1624 & 0.0385  \\
bottleneck units                & 230      & 580    & 270     \\
optimizer                       & adam     & adamax  & adam   \\
activation function             & selu & selu   & tanh    \\
$\lambda_{c}$                  & 0        & 0      & 0       \\
$\alpha_{c}$                  & 0.0001        & 0      & 0 \\
learning rate                   & 0.01    & 0.01  & 0.0001  \\
training epoch                  & 60       & 80     & 100     \\
mini-batch size                 & 55       & 40    & 80     \\ \hline
\end{tabular}
\end{table}

\begin{table}[!h]
\centering
\caption{Parameter configurations of the autoencoder that integrates data}
\begin{tabular}{cccc}
\hline
\multirow{2}{*}{hyperparameter} & \multicolumn{3}{c}{value}   \\
                                & GBM      & OV     & BRCA    \\ \hline
dropout rate                    & 0.1071   & 0.1515 & 0.0647  \\
bottleneck units                & 400      & 365    & 590     \\
optimizer                       & adamax   & adam  & rmsprop   \\
activation function             & softsign & softsign   & tanh    \\
$\lambda_\mathrm{con}$                  & 0        & 0      & 0       \\
$\alpha_\mathrm{con}$                  & 0.00001  & 0      & 0 \\
learning rate                   & 0.01    & 0.01  & 0.001  \\
training epoch                  & 40       & 60     & 80     \\
mini-batch size                 & 40       & 80    & 120     \\ \hline
\end{tabular}
\end{table}

\begin{table}[!h]
\centering
\caption{Parameter configurations of the risk prediction network}
\begin{tabular}{cccc}
\hline
\multirow{2}{*}{hyperparameter} & \multicolumn{3}{c}{value}   \\
                                & GBM      & OV     & BRCA    \\ \hline
dropout rate 1                   & 0.3307   & 0.1753 & 0.1294  \\
dropout rate 2                   & 0.3307   & 0.4039 & 0.2553  \\
bottleneck units 1               & 60      & 195    & 315     \\
bottleneck units 2               & 60      & 100    & 435     \\
optimizer                       & nadam   & rmsprop  & nadam   \\
activation function             & sigmoid & tanh   & sigmoid    \\
$\lambda_{r}$                  & 0.00001  & 0.01      & 0.0001  \\
$\alpha_{r}$                  & 0       & 0.001        & 0 \\
learning rate                   & 0.001    & 0.0001  & 0.0001  \\
training epoch                  & 10       & 70     & 90     \\
mini-batch size                 & 50       & 30    & 500     \\ \hline
\end{tabular}
\end{table}

 \section*{Acknowledgments}
The corresponding author: Qiulian Fang (E-mail: qiulianfang@csu.edu.cn). The authors would like to thank Yuchen Li, for his advice through the process.

The results shown here are in whole or part based upon data generated by the TCGA Research Network: \url{https://www.cancer.gov/tcga}.

\ifCLASSOPTIONcaptionsoff
  \newpage
\fi



\bibliographystyle{IEEEtran}%
\bibliography{bare_adv}
%


%






\end{document}